\documentclass[12pt]{iopart}
\usepackage[isolatin]{inputenc}
\usepackage{graphicx}
\usepackage{textcomp}
\usepackage{siunitx}

\begin{document}

\title[Removal of GaAs growth substrates from II-VI semiconductor heterostructures]{Removal of GaAs growth substrates from II-VI semiconductor heterostructures}
\author{S. Bieker, P. Hartmann, T. Kie{\ss}ling, M. R\"uth, C. Schumacher, C. Gould, W. Ossau and L. W. Molenkamp}
\address{Physikalisches Institut (EP3) der Universität Würzburg, 97074 Würzburg, Germany}
\ead{sbieker@physik.uni-wuerzburg.de}

\begin{abstract}
We report on a process that enables the removal of II-VI semiconductor epilayers from their GaAs growth substrate and their subsequent transfer to arbitrary host environments. The technique combines mechanical lapping and layer selective chemical wet etching and is generally applicable to any II-VI layer stack. We demonstrate the non-invasiveness of the method by transferring an all-II-VI magnetic resonant tunneling diode. High resolution X-ray diffraction proves that the crystal integrity of the heterostructure is preserved. Transport characterization confirms that the functionality of the device is maintained and even improved, which is ascribed to completely elastic strain relaxation of the tunnel barrier layer.

\end{abstract}
\maketitle

\section{Introduction}

Epitaxial growth techniques are a key technology for the fabrication of modern integrated electronics. A fundamental requirement for the growth of high crystalline quality epilayers is a proper match of the lattice constants of the grown heterostructures and the supporting substrate. This restriction often limits the choice of substrate materials, which in turn may prohibit the commercial applicability of certain device architectures or reduce performance. Epitaxial lift-off techniques (ELO) as proposed by Yablonovitch et al. \cite{Yablonovitch:1987bk} add a new degree of flexibility by enabling the removal of the resulting heterostructures from their growth substrate. ELO further allows for the transfer to new host environments, which may even enhance functionality. Several application ideas for ELO processes have been put forward, ranging from improved thermal management \cite{Das:2005gz} to cost-effective GaAs based photovoltaics and optoelectronics \cite{Yoon:2010bo}. 

The most widely used ELO technique relies on a sacrificial layer of Al$_x$Ga$_{1-x}$As that is grown between the GaAs substrate and the intended GaAs-based heterostructure. This technique exploits the large difference in etch rates between GaAs and Al$_x$Ga$_{1-x}$As in hydrofluoric acid (HF) \cite{Yablonovitch:1987bk}. Similar techniques have been developed for more specialized purposes, including laser lift-off for group III-nitride films \cite{Kelly:1996dq,Das:2005gz}. Only recently progress has been made in the development of lift-off routines for ZnSe-based heterostructures employing MgS \cite{Bradford:2005cg,Balocchi:2005ex} and ZnMgSSe release layers \cite{Moug:2009dn}. 

In this letter we demonstrate an ELO process that combines mechanical lapping and layer selective chemical wet etching. In contrast to previous work, this does not require a sacrificial release layer. The technique thereby simplifies the growth and is also applicable to heterostructures that were not meant to undergo lift-off when designed. A main achievement is the strongly reduced contact time with wet chemical agents, which minimizes negative effects on material quality in sensitive heterostructures. Our process is demonstrated for an all-II-VI compound semiconductor resonant tunneling diode (RTD), but is portable to other material systems. Current-voltage characteristics and X-ray diffraction are studied to assess the impact of the procedure on the structural integrity of the crystal. RTDs are used for this purpose as they are known to be highly sensitive to small changes in the material quality of the heterostructure.

\begin{figure}[tb]
	\centering
	\includegraphics[width=0.5\linewidth]{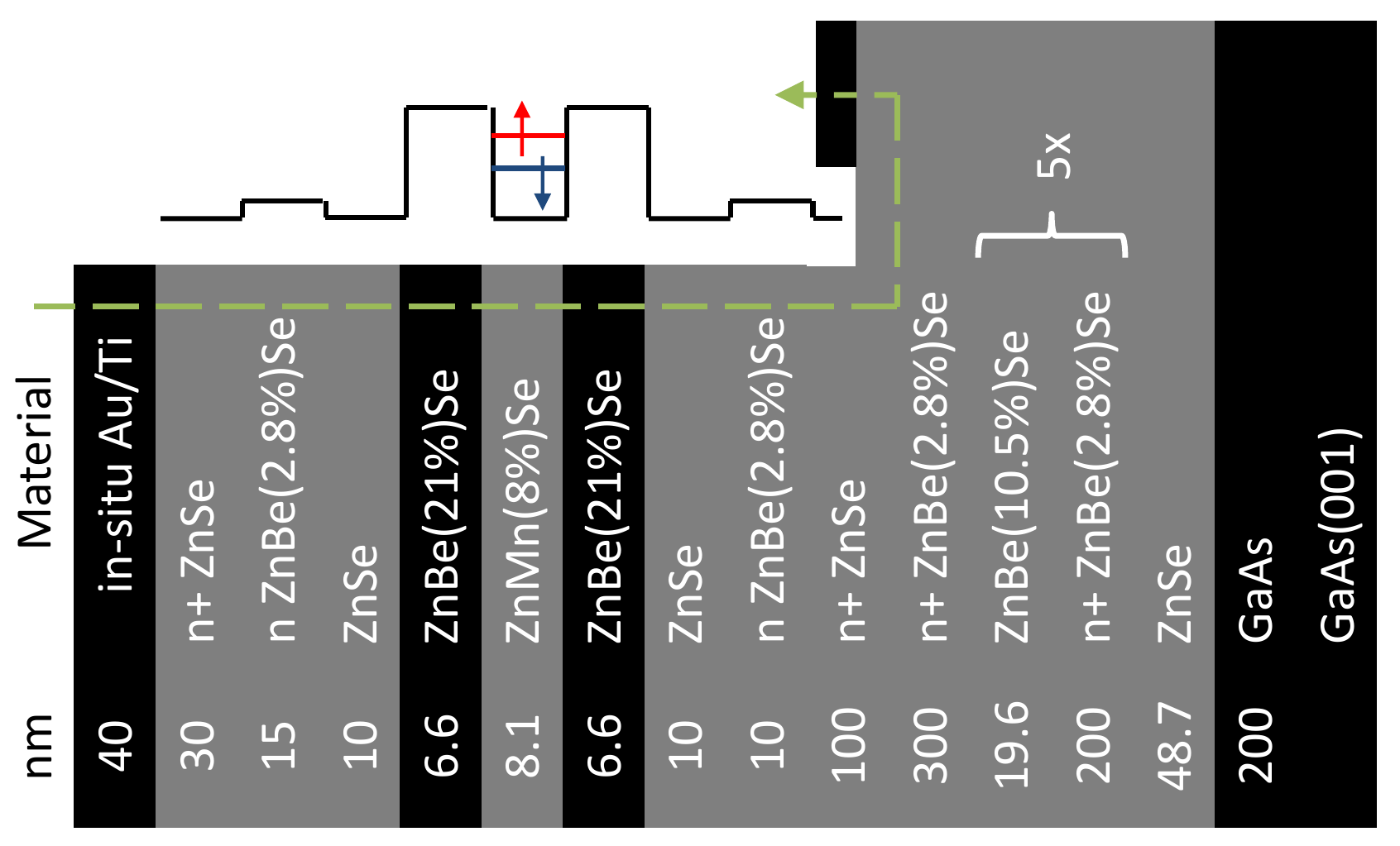}%
	\caption{Layer sequence and conduction band profile of our II-VI RTDs. The green arrow schematically indicates the current path through the pillar structure.}
	\label{Fig:layerstack}
\end{figure}

\section{The lift-off technique} 

The test structure RTDs are grown by molecular beam epitaxy (MBE) on (001)-oriented GaAs substrates. Its layer stack is shown in Fig.~\ref{Fig:layerstack} and a thorough discussion of the device and its layer properties can be found in \cite{Slobodskyy:2003jr}. The total thickness of the active device region, including its II-VI multilayer buffer, is roughly \SI{1}{\micro\metre}. We use a two-step process to detach the all-II-VI heterostructure from the GaAs substrate. In the first step, \SI{320}{\micro\metre} of GaAs are removed through mechanical lapping. A selective wet chemical etchant thereafter removes the remaining \SI{30}{\micro\metre} of substrate material.

The key ingredient of the method is the sample preparation prior to the lapping step. II-VI compound epilayers are mechanically softer than the GaAs substrate. Strong shear forces acting on the II-VI layers will therefore ruin the structural integrity and destroy the device. To decouple shear forces from the sensitive II-VI layers during the lapping process we developed a method that provides for a rigid silicon frame to protect the active device layers. This silicon frame is fabricated from a \SI[product-units=single]{10x10}{\milli\metre\squared} piece of silicon covered with a \SI{200}{\nano\metre} SiO$_2$ layer. Standard optical lithography is used to pattern a square window of \SI[product-units=single]{5x5}{\milli\metre\squared}. Within this window, the SiO$_2$ cap is removed by a buffered oxide etch of 7:1 NH$_4$F to HF. The remaining SiO$_2$ is then used as an etch mask in the last step, in which the entire silicon layer is inserted into KOH (\SI{20}{\percent}) at \SI{80}{\celsius}. Here the exposed silicon layer is etched from both sides. This produces a silicon frame of roughly half the thickness of the original layer with a \SI[product-units=single]{5x5}{\milli\metre\squared} opening.   

\begin{figure}[tb]
	\begin{minipage}[t]{0.05\linewidth}
			(a)
	\end{minipage}
	\begin{minipage}[t]{0.35\linewidth}
			\includegraphics[width=\linewidth]{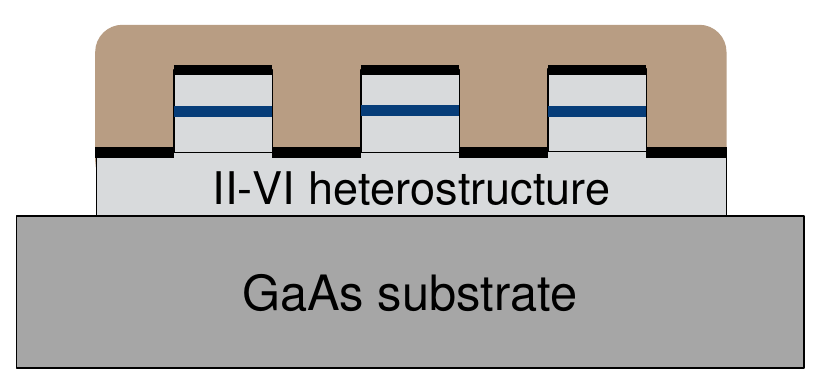}
	\end{minipage}
	\begin{minipage}[t]{0.05\linewidth}
			(b)
	\end{minipage}
	\begin{minipage}[t]{0.50\linewidth}
			\includegraphics[width=\linewidth]{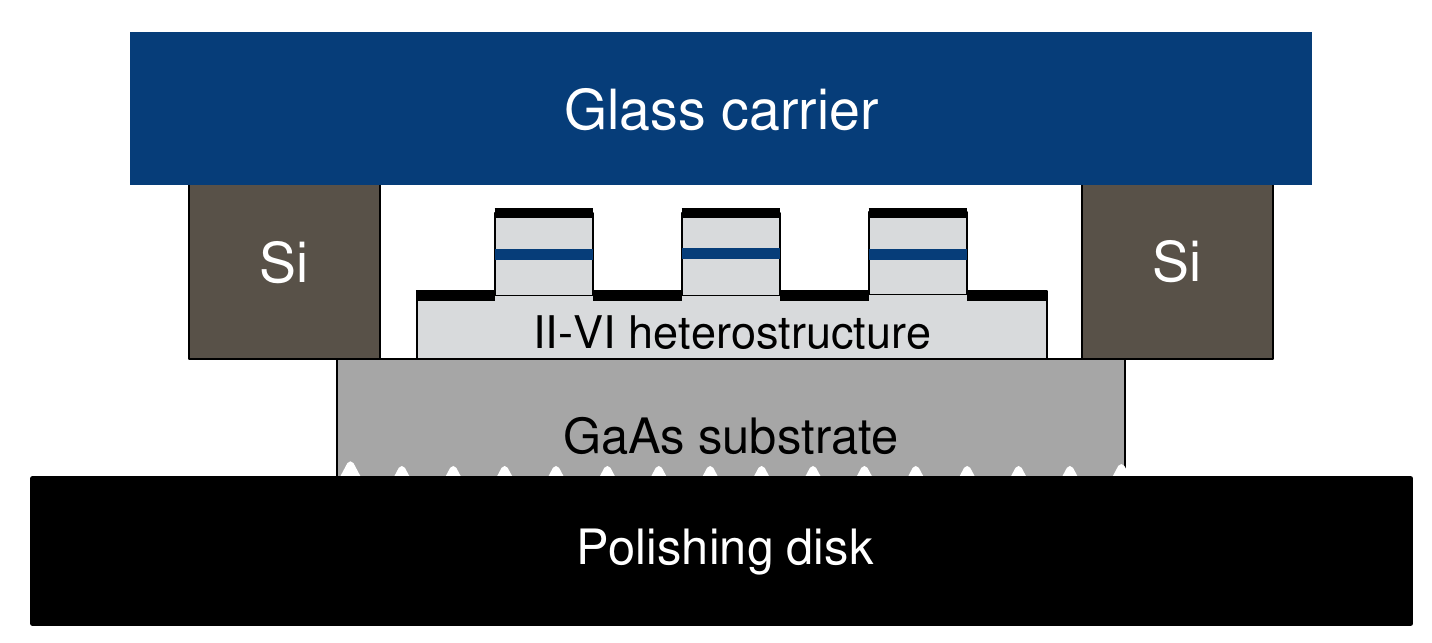}
	\end{minipage}
\caption{Schematic of the process described in the text. (a) Pre-patterned RTD pillars are covered with photo resist. The GaAs substrate around is uncovered using wet chemical etchants. (b) A silicon frame is attached to the bare substrate to decouple RTD pillars from shear forces during the mechanical lapping step.}
\label{Fig:process}
\end{figure}

Fig.~\ref{Fig:process} illustrates the subsequent steps of the process. We start out from pre-patterned RTD pillars as described in \cite{Slobodskyy:2003jr}. These stacks are not etched down completely to the GaAs substrate and reside on II-VI material. In the first step we define the sample areas to be detached and cover these with an organic photo resist. The size of the defined area is limited only by the dimensions of the silicon frame. On the remaining sample area we remove the metal layers. For this step we prepare a 1:1:1 mixture of thiourea, sodium sulfate and potassium ferricyanide (each in a \SI{10}{\percent} aqueous solution). The sample is then alternately dipped into the above etchant and a 1:200 aqueous solution of HF until the metal is completely gone. The uncovered II-VI layers are wet chemically etched down to the GaAs substrate by a solution of \SI{1}{\percent} bromine in ethylene glycole (see Fig.~\ref{Fig:process}a). The end of this etching step is reached as soon as the characteristic etalon fringes that stem from the II-VI layers disappear, leaving a free standing mesa of to-be-detached II-VI material on bare GaAs. These mesa are then placed inside the silicon frame, which is glued onto the GaAs using Apiezon wax. Finally, the silicon frame is attached to the rigid glass carrier of the polishing machine. This configuration leaves the sensitive RTD pillars inside the resulting cavity and no shear forces act on them during the subsequent mechanical lapping step (Fig. \ref{Fig:process}b).

During the MBE growth process the host substrate is often glued to the molybdenum substrate holders with indium adhesive to provide good thermal contact. Any such indium residues must be removed from the backside of the substrate as they would reduce the abrasive efficiency of the lapping step. This is done by immersion in HCl (\SI{37}{\percent}).

After this sample preparation, the glass carrier is attached to the underpressure jigface of the polishing machine. We use a commercial precision lapping and polishing machine (Logitech PM4) equipped with a rotating polishing disk that is continuously fed with calcined aluminum oxide powder (\SI{9}{\micro\metre} abrasive grain size) dissolved in distilled water. A micrometer gauge on the jig allows for control of the remaining sample thickness. As the first step of thinning, \SI{320}{\micro\metre} of GaAs substrate are removed through mechanical lapping. 

The thinned sample is then detached from the silicon frame by immersing the glass carrier in trichloroethylene at \SI{60}{\celsius}. The sample is cautiously removed from the solvent and glued topside down onto a clean glass plate. To selectively remove the residual GaAs substrate we use an 84:16 mixture of NaOH (\SI{5}{\percent}) and \mbox{H$_2$O$_2$} (\SI{31}{\percent}). Complete removal of the GaAs material is easily recognized by the appearance of the yellow II-VI layers underneath. Immersing the glass carrier again in trichloroethylene at \SI{60}{\celsius} detaches the II-VI layer stack from the carrier. The free floating lifted layer is finally picked up using the new host substrate of choice. If necessary, residual wax from previous process steps is removed with acetone, isopropyl and distilled water. After drying the sample film bonds to the new host substrate by van der Waals forces \cite{Yablonovitch:1987bk,Yablonovitch:1990bd} without the need for additional adhesive.

\begin{figure}[t]
	\centering
		\begin{minipage}{0.35\linewidth}
			 \includegraphics[width=\linewidth]{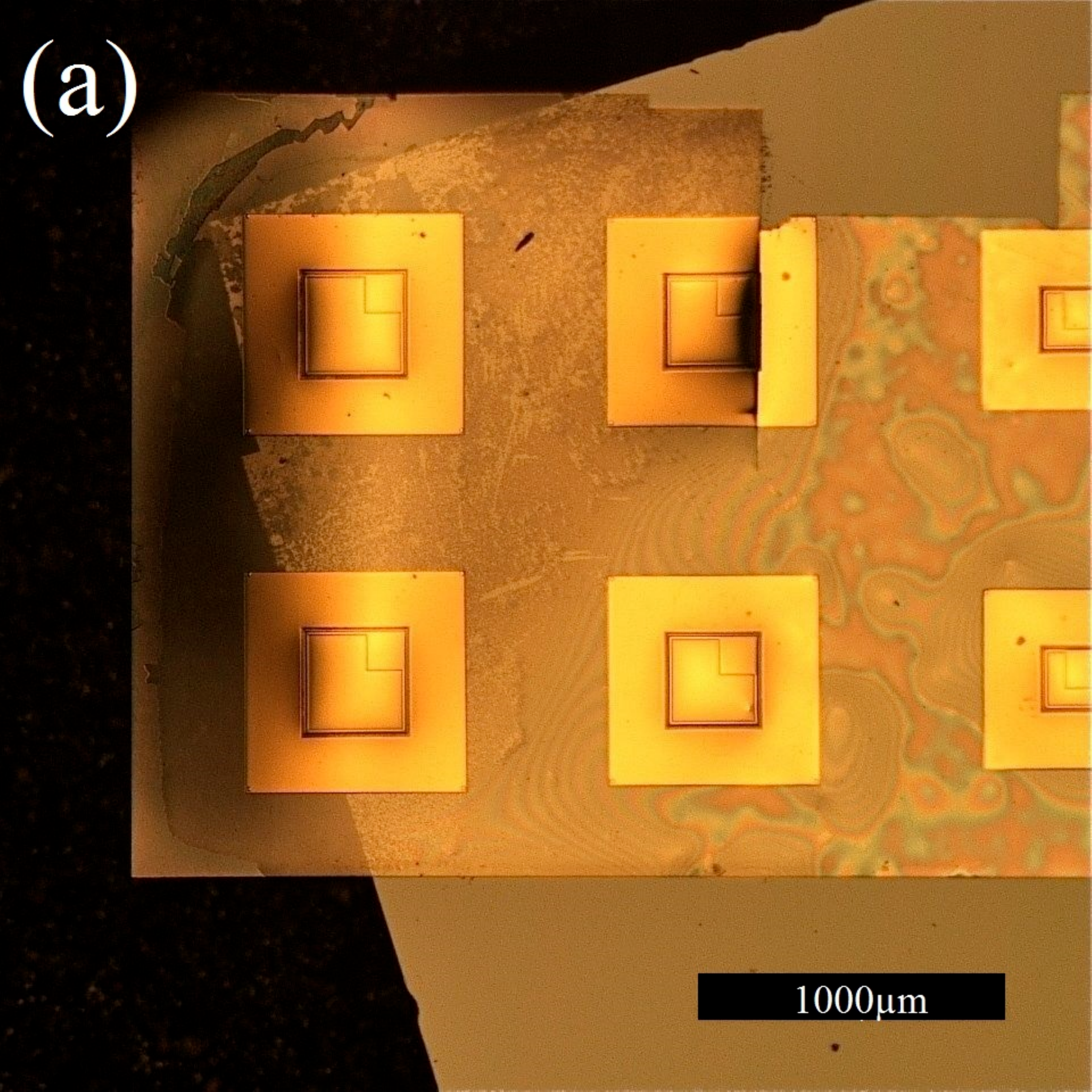}
		\end{minipage}
		\begin{minipage}{0.35\linewidth}
			 \includegraphics[width=\linewidth]{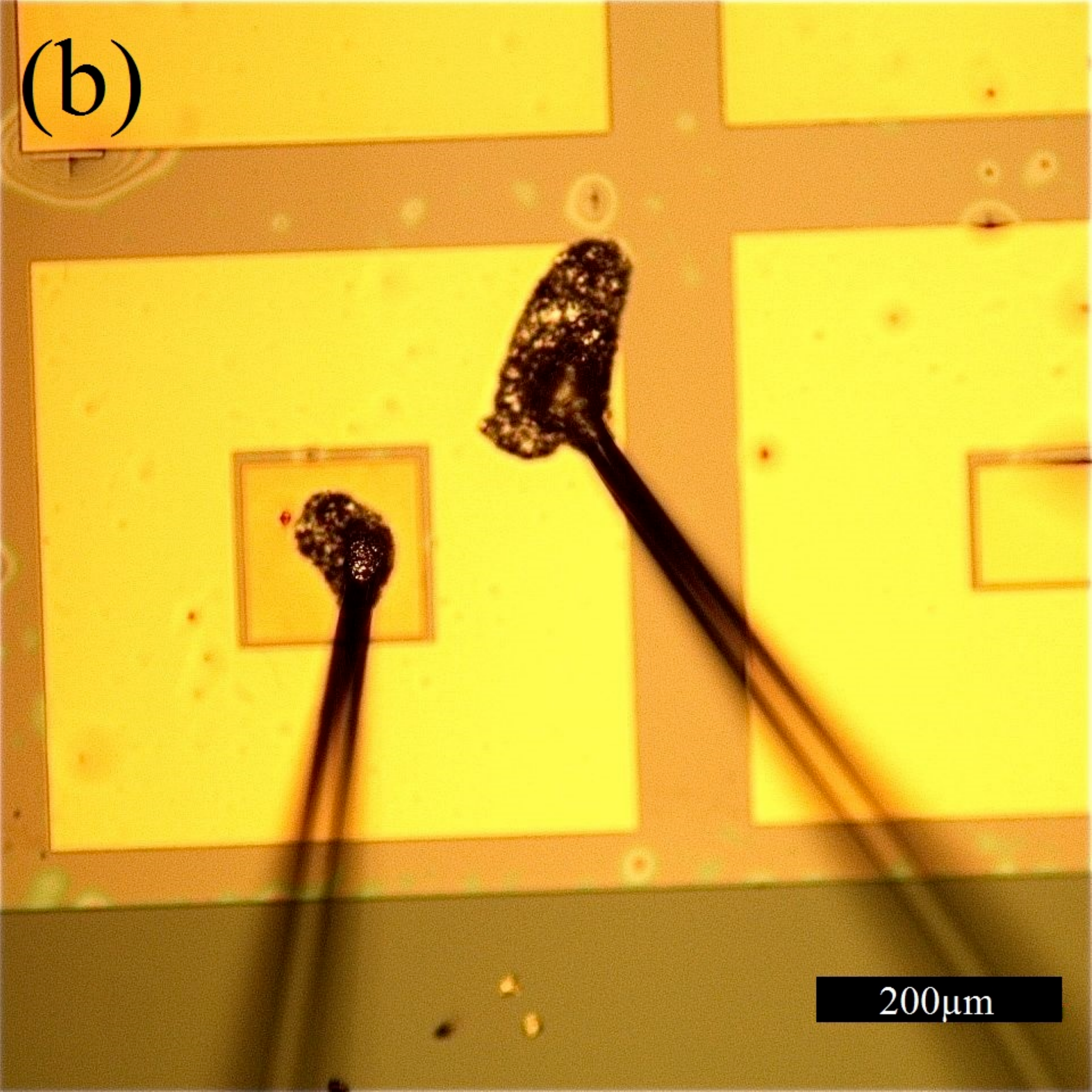}
		\end{minipage}
	\caption{(a) Sample film after removal of the GaAs as described in the text. The apparent stiffness of the lifted film indicates that the crystal integrity of the heterostructure is preserved. (b) One of the pre-patterned transport pillars is contacted with droplets of an electrically conductive epoxy after removal of the growth substrate.}
	\label{Fig:liftedsamples}
\end{figure}

Fig.~\ref{Fig:liftedsamples}a displays a photograph of a lifted layer that is  overlapping with the edge of the wafer that was used to capture it from the solvent. The obvious stiffness of the thin film is a strong indication that the crystalline integrity of the heterostructure is preserved during the process.

The thinned II-VI heterostructures do not withstand the stress from standard wire bonding. For wiring the device we therefore apply small droplets of a two component electrically conductive epoxy adhesive to the contact pads of the pillars. Fig.~\ref{Fig:liftedsamples}b shows a single RTD pillar from such a lifted layer, with gold wires attached to the top and backside contacts via conductive epoxy droplets.

\begin{figure}[tb]
\centering
  \includegraphics[width=0.7\linewidth]{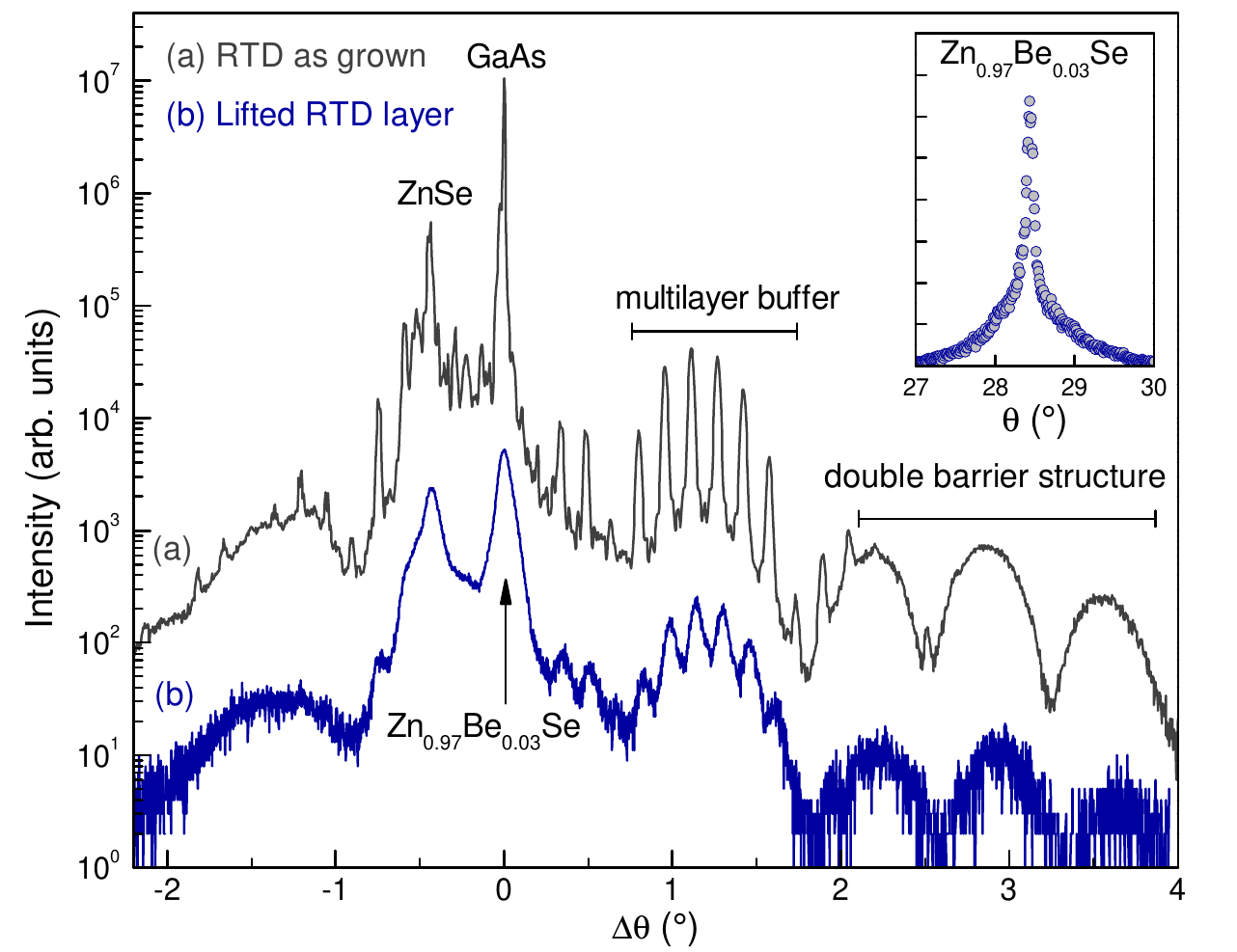}\\
  \caption{$\omega-2\theta$ HR-XRD scans of lifted and as-grown II-VI RTD structures. All distinct features are reproduced after removal of the substrate, demonstrating that our lift-off routine preserves the structural integrity of the crystal. Inset: $\omega$-scan of the Zn$_{0.97}$Be$_{0.03}$Se peak after substrate removal revealing a geometric broadening compared with the as-grown structures.}
  \label{fig:O2Tscan}
\end{figure}

\section{Sample characterization after lift-off}

We use high-resolution X-ray diffraction (HR-XRD) and current-voltage characterization to assess the impact of the lift-off process \cite{Frey:2010jv}. Fig.~\ref{fig:O2Tscan} shows a direct comparison of $\omega-2\theta$ scans for the same RTD sample before and after removal of the substrate. The sharp feature of the as-grown sample at $\Delta\theta =$ \ang{0} is the 004 reflection of the GaAs substrate. With the substrate removed, only the underlying reflection from the \SI{200}{\nano\metre} layer of lattice matched Zn$_{0.97}$Be$_{0.03}$Se remains. Due to its slightly larger lattice constant, the ZnSe 004 reflection is shifted to \ang{-0.45}. The distinctive diffraction pattern centered at \ang{1.2} stems from the II-VI multilayer buffer. The noticeable beating in the region from \SIrange[range-units = single]{2}{4}{\degree} results from X-ray interference between the symmetric Zn$_{0.79}$Be$_{0.21}$Se tunnel barriers. The $\omega$-scans of the ZnSe and Zn$_{0.97}$Be$_{0.03}$Se reflections reveal increased full width at half maximum (FWHM) values of \ang{;;433} and \ang{;;550} compared to typical values of \ang{;;20} on the as-grown structures (not shown). The $\omega$-width is a measure of the degree of tilt variation of the
lattice planes that contribute to a given x-ray reflex. Both the multilayer buffer features and the double barrier beating are, however, retained in the lifted layer. We therefore conclude that structural integrity of the crystal is preserved after lift-off, but that the layer does not rest perfectly flat on the new substrate.  

\begin{figure}[tb]
\centering
  \includegraphics[width=\linewidth]{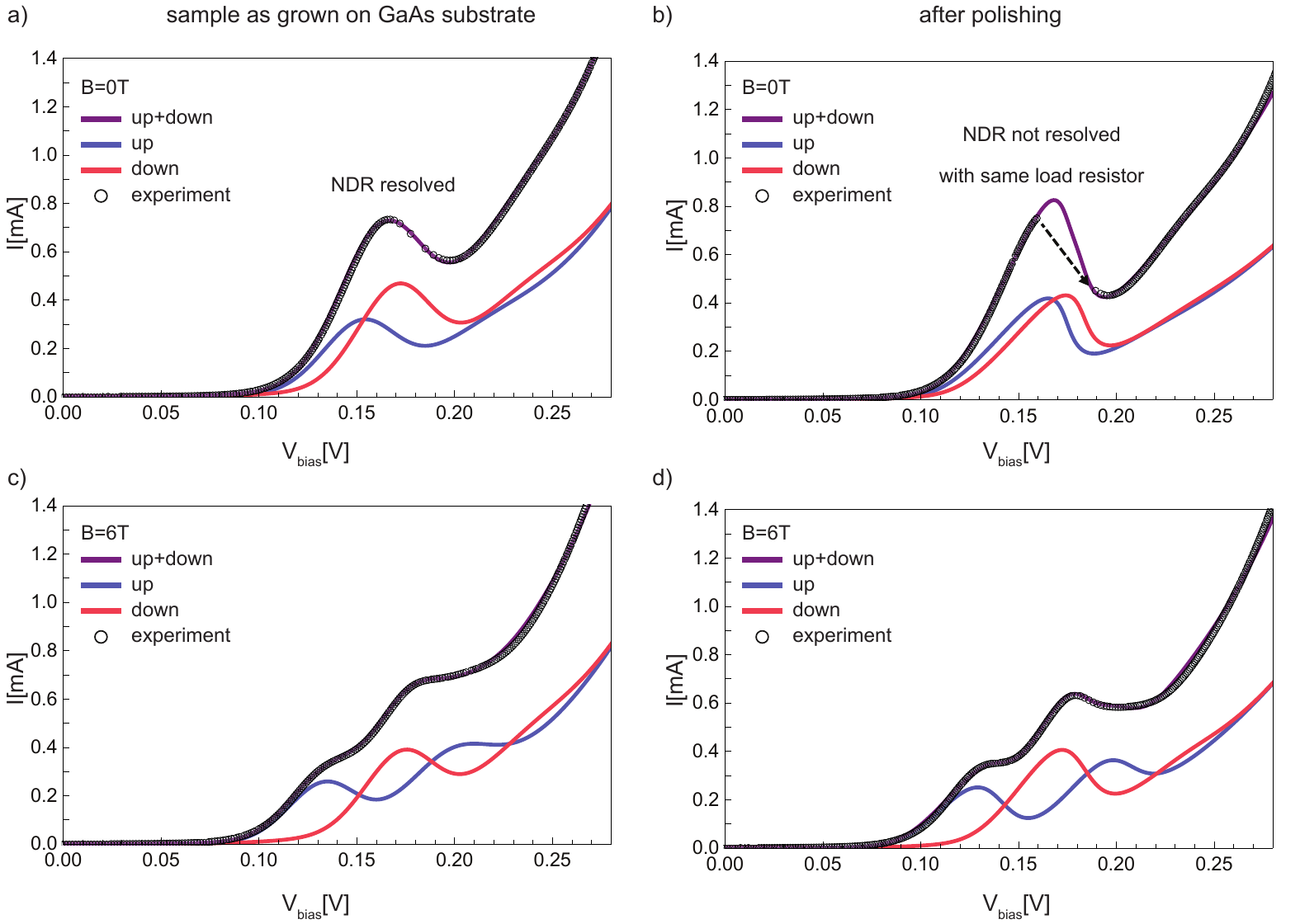}\\
  \caption{I-V characteristics of a dilute magnetic quantum well RTD, before and after removal of the GaAs substrate, at $B=\SI{0}{\tesla}$ and $B=\SI{6}{\tesla}$.}
  \label{fig:ivcurves}
\end{figure}

We chose RTDs as test devices for our substrate removal process because their I-V characteristics are very sensitive to the crystal quality of the heterostructure as well as to the smoothness of the interfaces. Figure~\ref{fig:ivcurves}a shows the I-V characteristics at $B=\SI{0}{\tesla}$ for the as-grown sample. A second sample of the same pillar size and processed from the same heterostructure is characterized after removal of the GaAs substrate, with results shown in Fig.~\ref{fig:ivcurves}b. Both samples are measured using the same measurement setup, and with a load resistor of $R=\SI{6}{\ohm}$. For the zero field curve of this second sample, the negative differential resistance (NDR) part of the I-V characteristic cannot be resolved, as indicated by the dashed arrow indicating a jump in the bias voltage. This results from a current bistability produced by the measurements loading line \cite{Foster:1989fs}. Given that the overall resistance of both devices is very similar, the load line analysis of the circuits are equivalent, and the apparent bistability can only be attributed to a sharpening of the peak and the corresponding increase in NDR after the resonance.

The I-V characteristics (black circles) are fitted using the model discussed in \cite{Ruth:2011fr}. The blue and red lines depict the contributions to the total current (purple line) of the spin-up and down channels, respectively. The fits to the I-V characteristics indicate the presence of a zero field splitting. In reference \cite{Ruth:2011fr} it is argued that this splitting, as well as the broadening of the zero field peak, result from rough interfaces at the quantum well interfaces. This allows the QW structure to break down into an ensemble of tunneling structures in parallel, each with a small area, and acting as a bound magnetic polaron (BMP) state. Since in each of these areas, the well properties differ slightly, this explains the broadening of the resonance, while the statistics describing the  magnetization of bound magnetic polarons accounts for a remanent zero field splitting, similar to that observed for dilute magnetic CdSe quantum dot systems \cite{Gould:2006io}.

\begin{figure}[tb]
\centering
	\includegraphics[width=0.5\linewidth]{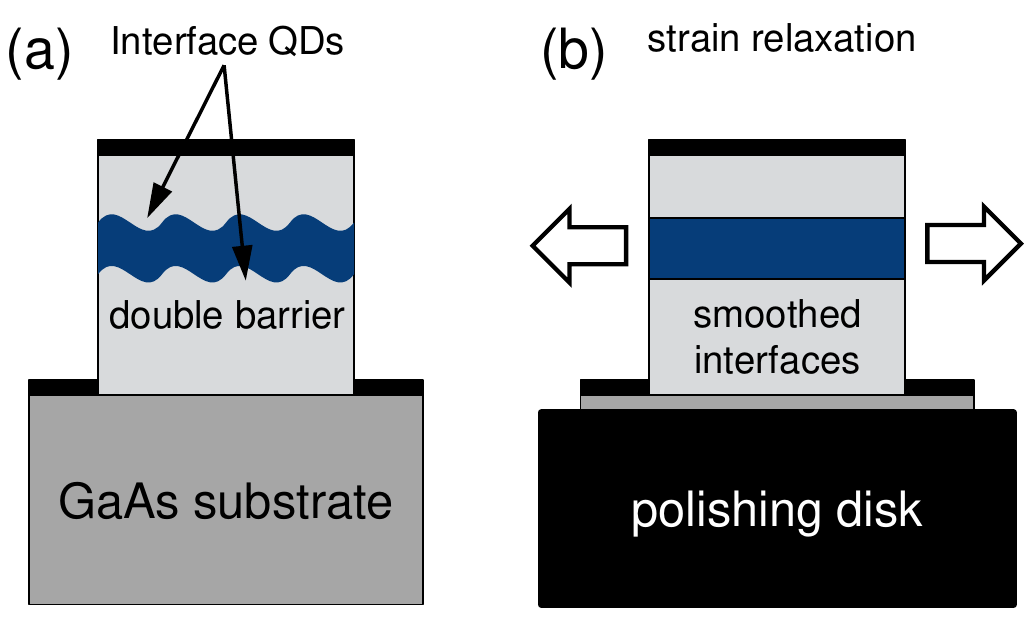}
	\caption{(a) Roughness of the quantum well interfaces allows for the formation of 0D type states which account for the remanent zero field splitting and the well width fluctuations that account for the broadening of the I-V resonance peak \cite{Ruth:2011fr}. (b) Removal of the GaAs substrate relaxes the strain imposed on the heterostructure, thus smoothening the interfaces and resolving the 0D type states.}
	\label{fig:strainrelaxation}
\end{figure}

The fits to the curves of Fig.~\ref{fig:ivcurves} give a peak broadening $\Gamma_0=\SI{8.9}{\milli\electronvolt}$ for the as-grown sample and one of $\Gamma_p=\SI{8.2}{\milli\electronvolt}$ for the lifted sample. This can be understood as the removal of the substrate relaxes the strain imposed on the II-VI layers by the imperfect lattice match to the substrate, and resolves the strain-induced imperfections at the interfaces as is schematically shown in Fig. \ref{fig:strainrelaxation}. The increased uniformity in QW thickness results in a sharpening of the resonance. This flattening of the interfaces also reduces the topographical interface features which sustain the formation of the BMP like states, and leads to a reduction of the zero field splitting, as is evident by comparing Fig.~\ref{fig:ivcurves}a and b.

\section{Summary and conclusions}

In summary, we have demonstrated a combined mechanical and chemical process that allows for the removal of GaAs growth substrates from II-VI heterostructures without the necessity of any sacrificial release layers. We have shown that the crystal integrity of even a fragile heterostructure is preserved during the substrate removal process, which is achieved by application of a silicon protection frame. Relaxation of strain previously induced by the lattice mismatch to the substrate even results in an increase in layer quality of the tested II-VI resonant tunneling diodes. The presented method can well be ported to a wide range of other material systems.

\section*{References}

\bibliography{References}

\begin{thebibliography}{10}

\bibitem{Yablonovitch:1987bk}
E~Yablonovitch, T~Gmitter, J~P Harbison, and R~Bhat.
\newblock {Extreme selectivity in the lift-off of epitaxial GaAs films}.
\newblock {\em Applied Physics Letters}, 51(26):2222, 1987.

\bibitem{Das:2005gz}
J~Das, W~Ruythooren, R~Vandersmissen, J~Derluyn, M~Germain, and G~Borghs.
\newblock {Substrate removal of AlGaN/GaN HEMTs using laser lift-off}.
\newblock {\em physica status solidi (c)}, 2(7):2655--2658, May 2005.

\bibitem{Yoon:2010bo}
J~Yoon, S~Jo, I~S Chun, I~Jung, H-S Kim, M~Meitl, E~Menard, X~Li, James~J
  Coleman, Ungyu Paik, and John~A Rogers.
\newblock {GaAs photovoltaics and optoelectronics using releasable multilayer
  epitaxial assemblies}.
\newblock {\em Nature}, 465(7296):329--333, May 2010.

\bibitem{Kelly:1996dq}
M~K Kelly, O~Ambacher, B~Dahlheimer, G~Groos, R~Dimitrov, H~Angerer, and
  M~Stutzmann.
\newblock {Optical patterning of GaN films}.
\newblock {\em Applied Physics Letters}, 69(12):1749, 1996.

\bibitem{Bradford:2005cg}
C~Bradford, A~Currran, A~Balocchi, B~C Cavenett, K~A Prior, and R~J Warburton.
\newblock {Epitaxial lift-off of MBE grown II--VI heterostructures using a
  novel MgS release layer}.
\newblock {\em Journal of Crystal Growth}, 278(1-4):325--328, May 2005.

\bibitem{Balocchi:2005ex}
A~Balocchi, A~Curran, T~C~M Graham, C~Bradford, K~A Prior, and R~J Warburton.
\newblock {Epitaxial liftoff of ZnSe-based heterostructures using a II-VI
  release layer}.
\newblock {\em Applied Physics Letters}, 86(1):011915, 2005.

\bibitem{Moug:2009dn}
R~Moug, C~Bradford, A~Curran, F~Izdebski, I~Davidson, K~A Prior, and R~J
  Warburton.
\newblock {Development of an epitaxial lift-off technology for II--VI
  nanostructures using ZnMgSSe alloys}.
\newblock {\em Microelectronics Journal}, 40(3):530--532, March 2009.

\bibitem{Slobodskyy:2003jr}
A~Slobodskyy, C~Gould, T~Slobodskyy, C~Becker, G~Schmidt, and L~Molenkamp.
\newblock {Voltage-Controlled Spin Selection in a Magnetic Resonant Tunneling
  Diode}.
\newblock {\em Physical Review Letters}, 90(24):246601, June 2003.

\bibitem{Yablonovitch:1990bd}
E~Yablonovitch, D~M Hwang, T~J Gmitter, L~T Florez, and J~P Harbison.
\newblock {Van der Waals bonding of GaAs epitaxial liftoff films onto arbitrary
  substrates}.
\newblock {\em Applied Physics Letters}, 56(24):2419--2421, 1990.

\bibitem{Frey:2010jv}
A~Frey, M~R{\"u}th, R~G Dengel, C~Schumacher, C~Gould, G~Schmidt, K~Brunner,
  and L~W Molenkamp.
\newblock {Semimagnetic II--VI semiconductor resonant tunneling diodes
  characterized by high-resolution X-ray diffraction}.
\newblock {\em Journal of Crystal Growth}, 312(7):1036--1039, March 2010.

\bibitem{Foster:1989fs}
T~Foster, M~Leadbeater, L~Eaves, M~Henini, O~Hughes, C~Payling, F~Sheard,
  P~Simmonds, G~Toombs, G~Hill, and M~Pate.
\newblock {Current bistability in double-barrier resonant-tunneling devices}.
\newblock {\em Physical Review B}, 39(9):6205--6207, March 1989.

\bibitem{Ruth:2011fr}
M~R{\"u}th, C~Gould, and L~W Molenkamp.
\newblock {Zero field spin polarization in a two-dimensional paramagnetic
  resonant tunneling diode}.
\newblock {\em Physical Review B}, 83(15):155408, April 2011.

\bibitem{Gould:2006io}
C~Gould, A~Slobodskyy, D~Supp, T~Slobodskyy, P~Grabs, P~Hawrylak, F~Qu,
  G~Schmidt, and L~Molenkamp.
\newblock {Remanent Zero Field Spin Splitting of Self-Assembled Quantum Dots in
  a Paramagnetic Host}.
\newblock {\em Physical Review Letters}, 97(1):017202, July 2006.

\end{thebibliography}

\end{document}